# Farey tree and devil's staircase of frequency-locked breathers in ultrafast lasers

Xiuqi Wu[1], Ying Zhang[1], Junsong Peng[1,2*], Sonia Boscolo[3], Christophe Finot[4], Heping Zeng[1,5,6*]

[1]State Key Laboratory of Precision Spectroscopy, East China Normal University, Shanghai 200062, China
[2]Collaborative Innovation Center of Extreme Optics, Shanxi University, Taiyuan, Shanxi 030006, China
[3]Aston Institute of Photonic Technologies, Aston University, Birmingham B4 7ET, United Kingdom
[4]Laboratoire Interdisciplinaire Carnot de Bourgogne, UMR 6303 CNRS – Université de Bourgogne Franche-Comté, F-21078 Dijon Cedex, France
[5]Chongqing Key Laboratory of Precision Optics, Chongqing Institute of East China Normal University, Chongqing 401120, China
[6]Shanghai Research Center for Quantum Sciences, Shanghai 201315, China

**Abstract** Nonlinear systems with two competing frequencies show locking or resonances. In lasers, the two interacting frequencies can be the cavity repetition rate and a frequency externally applied to the system. Conversely, the excitation of breather oscillations in lasers naturally triggers a second characteristic frequency in the system, therefore showing competition between the cavity repetition rate and the breathing frequency. Yet, the link between breathing solitons and frequency locking is missing. Here we demonstrate frequency locking at Farey fractions of a breather laser. The winding numbers show the hierarchy of the Farey tree and the structure of a devil's staircase. Numerical simulations of a discrete laser model confirm the experimental findings. The breather laser may therefore serve as a simple model system to explore universal synchronization dynamics of nonlinear systems. The locked breathing frequencies feature high signal-to-noise ratio and can give rise to dense radio-frequency combs, which are attractive for applications.

**Introduction**

Nonlinear systems with two competing frequencies show locking or resonances, in which the system locks into a resonant periodic response featuring a rational frequency ratio[1]. The locking increases with nonlinearity, and at subcritical values of the nonlinearity, the system has quasi-periodic responses between locked states, whilst the supercritical system may exhibit chaotic as well as periodic or quasi-periodic responses. A general feature of frequency locking is the robustness of the locked states to variations of system parameters, namely, the constancy of the frequency ratio (or winding number) over a range of parameters. Resonances have been investigated theoretically and experimentally in many physical systems including coupled oscillators[2], charge-density waves[3], Josephson junctions[4, 5] and the Van der Pol oscillator[6] amongst others[7], and their distribution in parameter space in the form of a devil's staircase[8] is well understood from the number theory concept of Farey trees[9, 10, 11, 12, 13, 14]. In optics, frequency-locking phenomena have been extensively studied in modulated semiconductor lasers, where an external frequency can be readily coupled to the nonlinear system by using a radio-frequency (RF) source[11, 15, 16, 17, 18], and the hierarchy of the Farey tree and structure of a devil's staircase can be rather easily observed when tuning the external frequency[11]. Frequency locking has also been demonstrated in other laser structures, such as fibre lasers with external loss modulation[19] or solid-state lasers operating in a two-mode regime[20]. Furthermore, although not explicitly mentioned by the authors, the subharmonic, harmonic and rational harmonic operation regimes of Kerr micro-resonators that were reported in Refs.[21, 22] imply a frequency-locking process. The generation of soliton molecules (i.e., stable bound states of two solitons) in a titanium-sapphire laser that was reported in Ref.[23] also evidences the occurrence of frequency locking: a subharmonic response of the soliton molecule was observed when the strength of the external driving force exceeded a certain threshold.

All the frequency-locking examples mentioned above relate to nonlinear systems where an external, accurately controllable modulation adds a new characteristic frequency to the system. Far less is experimentally known, by comparison, when the second frequency is not externally controlled and is intrinsic to the nonlinear system. This is particularly relevant to breathing solitons that have recently emerged as a ubiquitous mode-locked regime of ultrafast fibre lasers[24, 25, 26, 27, 28]. Breathing solitons, manifesting themselves as localised temporal / spatial structures that exhibit periodic oscillatory behavior, are found in various subfields of natural science, such as solid-state physics, fluid dynamics, plasma physics, chemistry, molecular biology and nonlinear optics[29]. Optical breathers were first studied experimentally in Kerr fibre cavities[30] and subsequently reported in optical micro-resonators[21, 31, 32]. They are currently attracting significant research interest in virtue of their connection with a range of important nonlinear dynamics, such as rogue wave formation[33, 34], the Fermi-Pasta-Ulam recurrence[35, 36, 37], turbulence[38], chimera states[39, 40], chaos[41] and modulation instability phenomena[42]. From a practical application perspective, breathers can increase the resolution of dual-comb spectroscopy[43] as the breathing frequency comes along additional tones in a frequency comb, and the breather regime in a laser oscillator can be used to generate high-amplitude ultrashort pulses without additional compressors[44, 45].

In this paper, we present the first in-depth study of the locking of breather oscillations to the cavity repetition frequency in a fibre laser. Besides the hurdle represented by the absence of an external driver to realize frequency locking, the excitation of breathing solitons in a fibre laser requires fine tuning of the laser parameters, where the breather mode-locking regime exists in a narrower parameter space than stationary mode locking[44]. Therefore, targeting frequency-locked breather states in the laser via trial and error is a laborious task. Here we show that such a difficulty can be circumvented by using an evolutionary algorithm (EA) based on the optimal parameter tuning of the intracavity nonlinear transfer function through computer-controlled polarisation control. Machine-learning strategies, referring to the use of statistical techniques and numerical algorithms to carry out tasks without explicit programmed and procedural instructions, are widely deployed in many areas of engineering and science[46]. In the field of ultrafast photonics, machine-learning approaches and the use of genetic and evolutionary algorithms have recently led to several dramatic improvements in dealing with the multivariable optimisation problem associated with reaching desired operating regimes in fibre lasers. In the present study, the merit function used in the EA optimisation procedure can distinguish between frequency-locked and unlocked breather states, thereby enabling fast and precise tuning of the laser to the target frequency-locked breather operation. The locked breather states show two unambiguous features: persistence under pump power and polarisation perturbations, and narrow linewidth and high signal-to-noise ratio (SNR) of the oscillation frequency in the electrical spectrum of the laser emission. Importantly, frequency-locked states occur in the sequence they appear in the Farey tree and within a pump-power interval given by the width of the corresponding step in the devil's staircase. This demonstrates that breather mode-locked fibre lasers exhibit the universal properties characteristic of nonlinear systems driven by two competing frequencies.

**Results**

**Frequency-locked and unlocked breathers in the laser**

To investigate the dynamics of breathers, we have built the fibre ring cavity that is sketched in Fig. 1(a). Pump light is provided by a laser diode operating at 980 nm and it is delivered to the unidirectional cavity through a wavelength-division multiplexer. A 1.25-m-long erbium-doped fibre segment constitutes the gain medium. Other fibres in the cavity are standard single-mode fibre from the pigtails of the optical components used, including an isolator and two collimators. The group-velocity dispersion (GVD) values of the two fibre types are 65, and –22.8 $ps^2$/km, respectively, yielding a normal net cavity dispersion of 0.009 $ps^2$ at the operating wavelength of ~1.5 $\mu$m. The repetition rate of the laser is $f_r$=34.2 MHz. The mode-locked laser operation is obtained thanks to an effective saturable absorber based on the nonlinear polarisation evolution (NPE) effect[47]. The nonlinear transfer function of the NPE-based mode locking is controlled by three wave plates based on liquid crystal (LC) phase retarders working together with a polarisation beam splitter (PBS). The PBS is also used as an output coupler. The emitted light from the laser is monitored by several diagnostic systems. A fraction is directly detected by a fast photodiode (PD1, Finisar XPDV2320R; 20-ps response time, 50-GHz bandwidth) plugged to a real-time oscilloscope (Agilent; 33-GHz bandwidth, 80-GSa/s sampling rate). The remaining laser output is sent through a time-stretch dispersive Fourier transform (DFT) setup consisting of a long segment of normally dispersive fibre that cumulates a group-velocity dispersion large enough for the stretched waveform to represent the spectral intensity of the initial pulse waveform[48]. From the photodetection of the DFT output signal on a fast photodiode (PD2), the optical spectrum for each pulse is obtained directly on the oscilloscope. Additional measurement devices are used to characterise the spectral properties of the laser output: an optical spectrum analyzer, an electrical spectrum analyzer (ESA) and a cymometer.

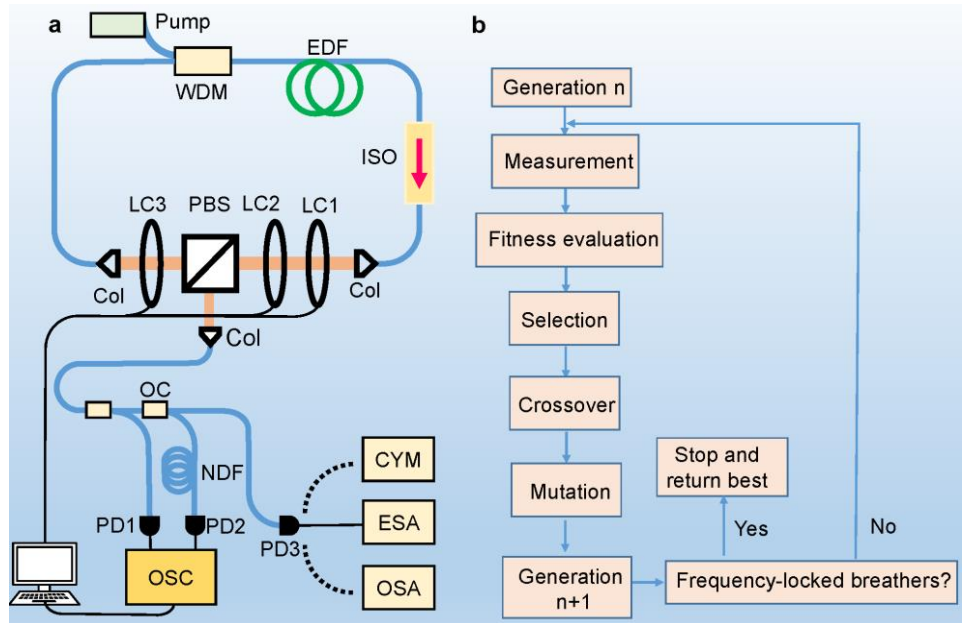

Fig. 1 (a) Experimental setup of the breather fibre laser. WDM: wavelength-division multiplexer; EDF: erbium-doped fibre; ISO: isolator; LC: liquid crystal phase retarder; PBS: polarisation beam splitter; Col: collimator; OC: optical coupler; NDF: normally dispersive single-mode fibre involved in the DFT measurements; PD: photodetector; OSC: real-time oscilloscope; CYM: cymometer; ESA: electronic spectrum analyzer; OSA: optical spectrum analyzer. (b) Illustration of the EA principle.

Breathing solitons can be excited in the laser cavity by tuning the gain (pump strength) and the cavity loss (polarisation controllers)[24]. Panels (a) and (b) of Fig. 2 show an example of a breather operation of the laser recorded at a pump power of 74 mW. In sharp contrast with soliton pulse shaping which generates uniform pulse trains, the train of output pulses shows periodic variations in intensity occurring, in the example of Fig. 2(a), across a well-defined period of 50 cavity roundtrips. Note that while Fig. 2(a) shows the photo-detected signal after time stretching, the same periodic evolution is also observed for the pulse train directly detected at the laser output. The corresponding spatio-spectral representation of the laser regime (Fig. 2(b)) evidences a periodic compression and stretching of the optical spectrum over cavity roundtrips, accompanied by synchronous periodic changes in pulse energy (white curve), which is a distinctive feature of breathing solitons. Variations in the system parameters may give rise to a different breather state in the laser as shown in panels (c) and (d) of Fig. 2, where the pump power is decreased to 73 mW: whilst the period of oscillation seems to be unchanged, the quality of the periodic behavior is clearly degraded in comparison with the previous case. The RF spectra of the laser emission taken from the ESA (Fig. 3) reveal the major difference between the two types of breather states. The breathing frequency of the unstable breather state shown in Fig. 2(c) exhibits a noisy and broad structure (Fig. 3(c, d)). By contrast, the stable breather state of Fig. 2(a) features a neat breathing frequency with narrow linewidth (0.5 Hz; see Supplementary Fig. 1 for details of the measurement) and high SNR (Fig. 3(a, b)). The measurements taken with the cymometer confirm the different stability properties of the breathing frequency for the two states (Fig. 3(e)). The breathing frequency of the stable breather state is $f_b$ =6.84 MHz exactly equalling one fifth of the fundamental repetition frequency, hence corresponding to a rational winding number of $f_b/f_r = 1/5$. As discussed later in this paper, this locked breathing frequency remains unchanged over a range of pump power values.

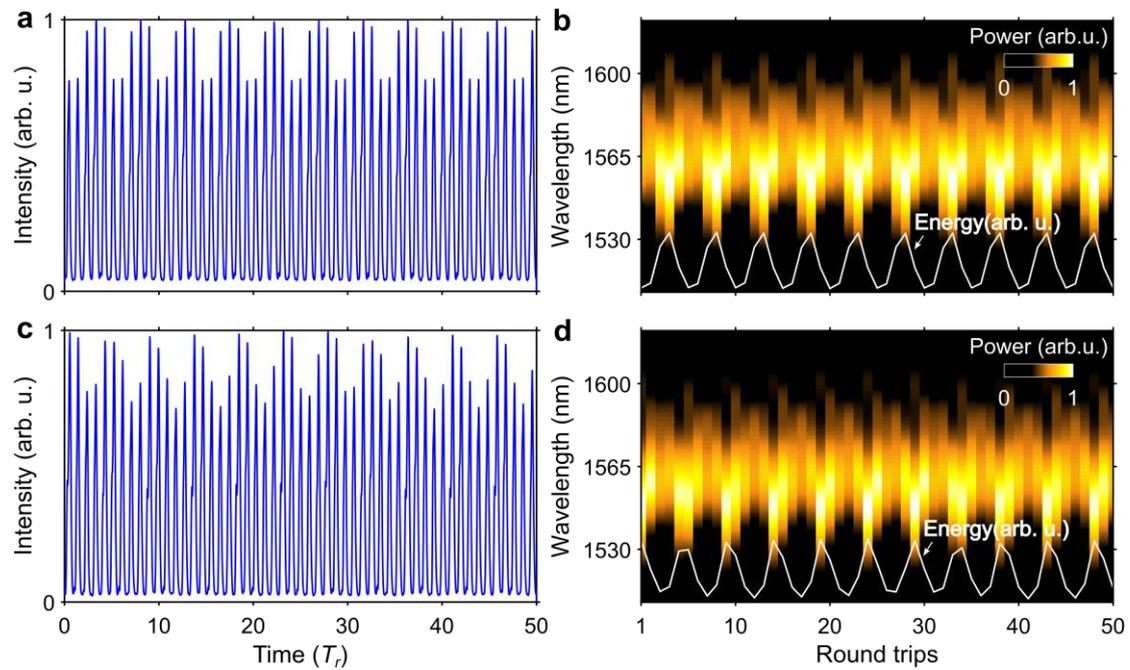

Fig. 2 Two different breather operations of the laser observed over 50 cavity roundtrips: (a, b) frequency-locked breather state showing a well-defined periodicity, and (c, d) frequency-unlocked breather state featuring degraded periodicity. Panels (a, c) show the photo-detected DFT output signals ($T_r$ is the roundtrip time), and panels (b, d) are the corresponding DFT recordings of single-shot spectra. The white curves in (b, d) represent the energy evolutions.

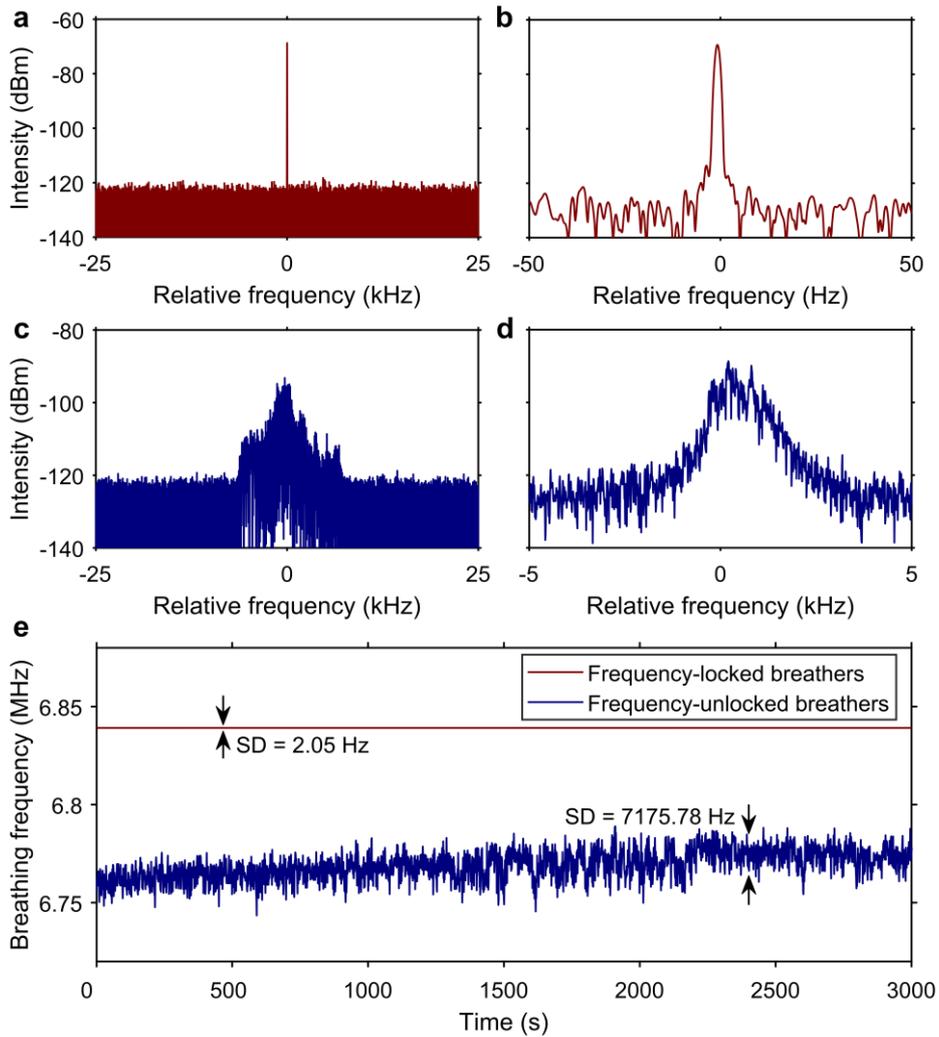

Fig. 3 RF spectral measurements of the breather states shown in Fig. 2. The reference frequency is one fifth of the fundamental repetition frequency. (a, b) Single-mode oscillation of the breathing frequency when frequency locking occurs measured over spans of 50 kHz and 100 Hz, respectively. (c, d) Unstable multimode oscillation of the breathing frequency measured over 50-kHz and 10-kHz spans. (e) Change in breathing frequency over time for the locked (red) and unlocked (blue) breather states, as measured with a cymometer. The standard deviation (SD) of the breathing frequency values is 2.05 Hz for the frequency-locked state and is 7175.78 Hz for the unlocked state.

**Evolutionary algorithm optimisation of frequency-locked breathers**

Reaching a frequency-locked breather state in our laser depends on precisely adjusting four parameters: the pump strength and three polarisation controllers, which is quite difficult to do manually. In Ref.[49], we have introduced an approach based on an EA for the search and optimisation of the breather mode-locking regime in ultrafast fibre lasers, which relies on specific features of the RF spectrum of the breather laser output. In the self-tuning regime, the operation state of the laser is characterised in real time with the oscilloscope, which is connected to a computer running the EA and controlling the polarisation state through the voltages applied on the LCs via a driver to lock the system to the desired breather regime (Figs. 1(a) and (b)). Yet, the merit function of the breather mode locking used in Ref.[49] is unable to distinguish between frequency-locked and unlocked breather states, where it usually breeds unlocked (unstable) states which have a wider parameter space. Here, we further develop our approach to directly pinpoint frequency-locked breathers so that the EA tunes the laser to these states only. To this end, we define a new merit function which takes into account the distinguishing trait of frequency-locked breather states, namely, a high SNR of the breathing frequency as shown in Fig. 3(a, b). The new merit function is given in Eq. (2) in the "Methods" section. An example of an optimisation curve (referring to a breather state with a winding number of 1/5) is presented in Fig. 4(a), which shows the evolution of the best and average merit scores of the population, as defined by Eq. (2), for each generation along with the corresponding evolution of the SNR of the breathing frequency. We can see that the SNR quickly increases and converges to a maximum value after 8 generations, thus indicating the establishment of a frequency-locked operation mode of the laser. The best merit score features a similar evolution. The measurements of the breathing frequency under pump power and ploarisation tuning shown in Figs. 4(b) and 4(c), respectively, confirm the operation of the laser in the target mode. The reliability of the merit function of the frequency-locked breather regime has been assessed by repeating the optimisation procedure numerous times, with the results showing that each time the SNR of the breathing frequency is high frequency locking occurs. Additional examples of optimisation curves (for breather states with the winding numbers 1/5 and 2/9) are given in Supplementary Fig. 2.

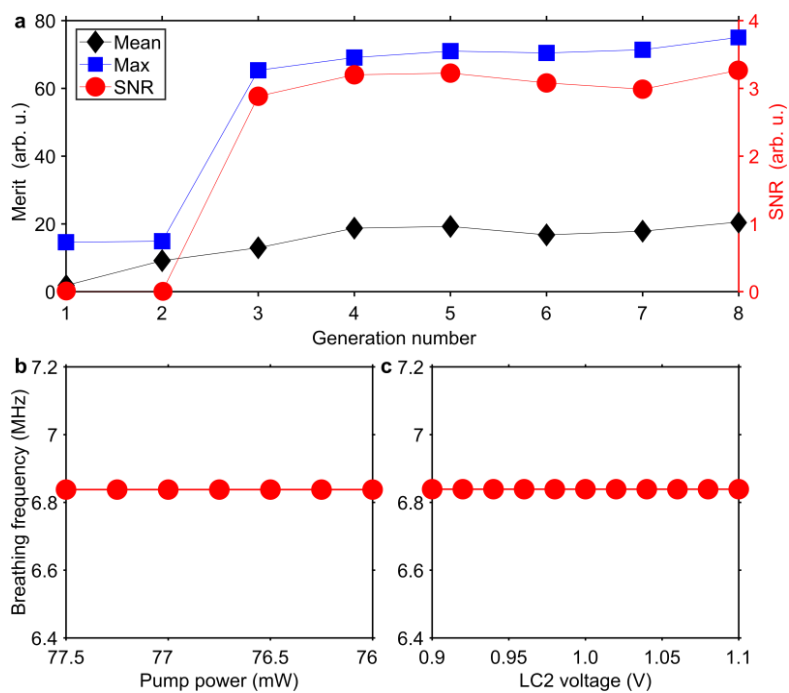

Fig. 4 (a) Evolution of the average (black diamonds) and maximum (blue squares) merit scores over successive generations, for the merit function given in Eq. (2) (Methods). Also shown is the corresponding evolution of the SNR of the breathing frequency (red circles). Persistence of the optimal state with variation of the (b) pump power and (c) polarisation (varied by changing the voltage on LC2).

**Farey tree and devil's staircase of the breather laser**
Benefiting from a reliable and efficient EA-based optimisation approach, we have explored the transitions between the different breather states of the laser that can be accessed by varying the pump power starting from the range corresponding to a 1/5 frequency-locked state. Figure

5(a) shows an example of a plot of the breathing frequency as a function of the pump power, revealing the presence of various plateaux (steps). The spectral measurements carried out with the ESA allow us to unambiguously relate the breathing frequencies associated with the plateaux to rational winding numbers: as shown in panels (b-d) of Fig. 5, when the laser operates in a frequency-locked state, the RF spectrum features a finite number $n$ of spectral lines below the cavity repetition frequency $f_r$ and equally spaced by $f_r/n$. For example, in panel (d) the frequency-locked breather regime brings about the excitation of a RF comb that is 41 times denser than that obtained when the laser operates in the usual single-pulse stationary regime. The most intense line in the spectrum is the breathing frequency $f_b$, and if this is the $m$th line from the short-frequency side, then the corresponding winding number is given by $m/n$. The temporal and spectral dynamics of the breather patterns belonging to the winding numbers 2/9 and 9/41 are given in the Supplementary Fig. 3.

Importantly, in Fig. 5(a) the winding numbers appear from left to right in the order predicted by the Farey tree, as shown in the inset of the figure, and the width of the step associated with a $m/n$ frequency-locked state depends on the level where $m/n$ appears in the Farey tree's hierarchy. The gaps (in pump power) between the stairs refer to quasi-periodic breather oscillations similar to the example shown in Fig. 2(c,d) and Fig. 3(c,d). The fractal dimension $D$ of the set of gaps can be extracted from the width of the steps (see Methods), and is calculated to be $D$ = 0.906±0.025, which is close to the value of 0.87 expected from a complete devil's staircase[9]. Note that fractal dimensions of 0.890±0.001 and 0.91±0.03 were reported in Refs.[11] and[8], respectively. Here, the small deviation (4%) from 0.87 partly results from the minimum power increment of the pump laser diode (0.1 mW). The fact that the steps associated with the winding numbers 7/32 and 9/41 consist of only one point in Fig. 5(a) is also due to this limitation, thus stressing the need for a very robust control of the system's properties. The process of formation of the devil's staircase is reversible: by decreasing the pump power, nearly the same staircase can be observed. We emphasise that contrarily to modulated external-cavity semiconductor lasers where the modulation frequency can be arbitrarily set hence the frequency-locked states expected according to the Farey tree can be easily accessed[11], in a breathing-soliton mode-locked laser the breathing frequency is established once the laser is fabricated, while it can be entrained by tuning the laser parameters. Nevertheless, the Farey tree and devil's staircase can still be observed, indicating the universal nature of this nonlinear system. Setting the laser to a slightly different initial polarisation state, Farey fractions belonging to other two parts of the Farey tree can be identified through the RF spectra while tuning the pump power (see Supplementary Figs. 4 and 5). In both cases, the calculated dimension of the set complementary to the stairs approaches that of a complete devil's staircase.

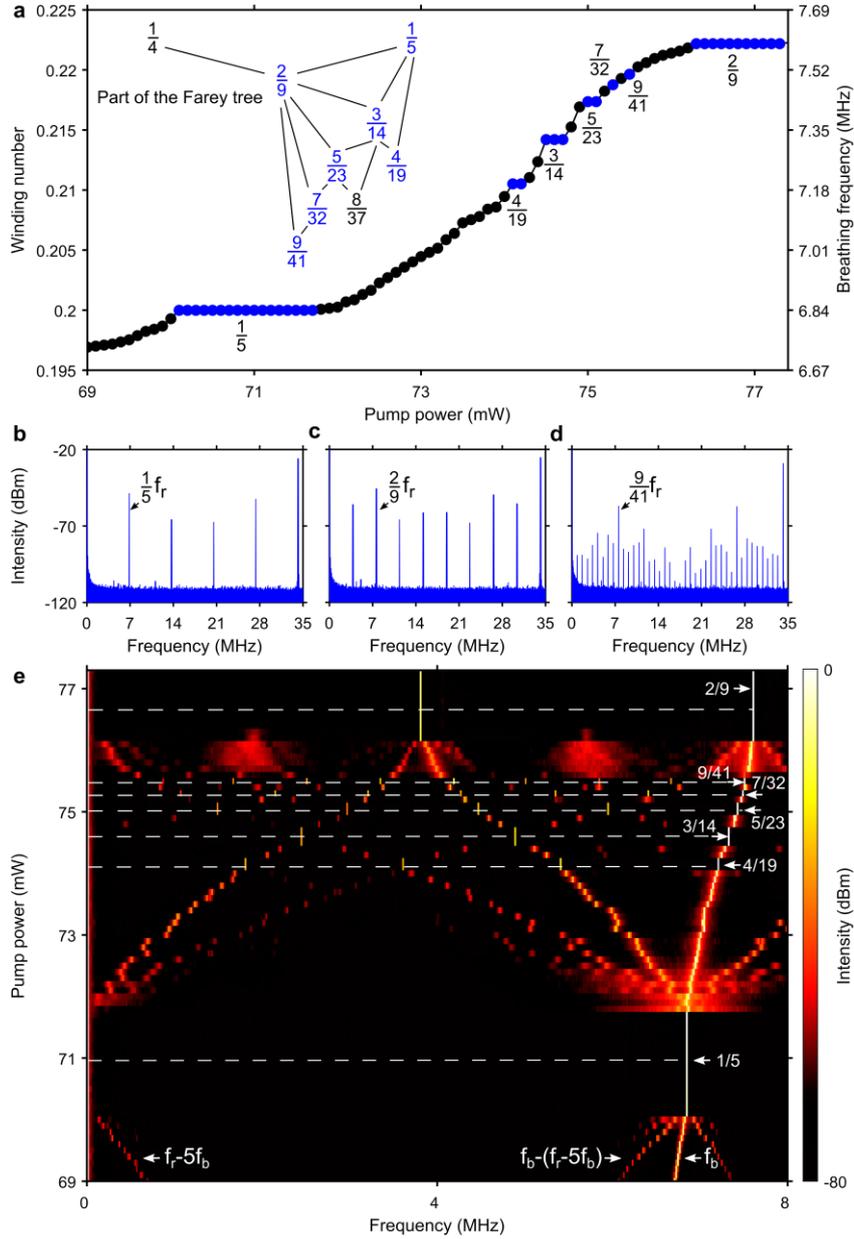

Fig. 5. RF spectra, Farey tree and devil's staircase. (a) Measured breathing frequency (winding number) as a function of the pump power. In the inset is shown the part of the Farey tree containing the observed Farey fractions. (b, c, d) RF spectra measured with the ESA showing dense frequency combs for the frequency-locked states corresponding to the winding numbers 1/5, 2/9 and 9/41, respectively. A new set of equidistant spectral lines fills in the frequency interval corresponding to the cavity repetition rate $f_r$ (34.2 MHz). (e) Map of spectral intensity in the space of radiofrequency and pump power, showing the build-up of rational winding numbers.

Figure 5(e) illustrates the build-up phase of frequency locking. Starting from a pump power of 69 mW, three radiofrequencies are present, namely the breathing frequency ($f_b$), the difference frequency between $f_r$ and 5th harmonic of $f_b$ ($f_r - 5f_b$), and the difference frequency between the first two ($6f_b - f_r$). As $f_r - 5f_b$ approaches zero, frequency locking occurs at the winding number 1/5. This winding number then experiences redshifts under pump power increments, generating other winding numbers. The map shown in Fig. 5(e) also evidences the very different spectral features of frequency-locked and quasi-periodic breather states. Therefore, even though the winding numbers 7/32 and 9/41 display only one point in Fig. 5(a), they can clearly be identified in this map, which reveals a richness of detail that has been largely overlooked in previous studies due to lack of high-quality RF spectral measurements. It is also noteworthy that changing the pump power by only 10% is enough to find seven frequency-locked states for the laser, whose power-stability properties are dictated by a devil's staircase. As a further note, we would like to emphasize that the frequency-locked states observed are reproducible but not self-starting, meaning that if the pump power is turned off when the laser operates in a locked state and then it is turned back on again, the laser does not return to that state instantaneously. To restore the frequency-locked operation, one can run the EA controlling the polarisation states again, which will quickly reset the laser to the desired state. Many such experimental tests have confirmed the reproducibility of the locked states.

To validate our experimental findings, we have performed numerical simulations of the laser using a scalar-field, lumped model that includes the dominant physical effects of the system on the evolution of a pulse over one round trip inside the cavity, namely, GVD and self-phase modulation for all the fibres, gain saturation and bandwidth-limited gain for the active fibre[50], and the discrete effects of a saturable absorber element (see 'Methods' section). The gain saturation energy in the model is related to the pump power in the experiment. Panels (a) and (b) of Fig. 6 show plots of the breathing frequency (winding number) as a function of the gain saturation energy when the latter is varied starting from the range corresponding to a 1/5 locked state with a step of 10 pJ and 1 pJ, respectively. With the smaller step, more plateaux are observed, thus confirming the fractal structure of the winding number distribution. It is seen in Fig. 6(b) that the model can reproduce the same part of the Farey tree from a breathing frequency of 1/5 to 2/9 as that observed in the experiment (Fig. 5(a)). The gaps (in gain saturation energy) between the stairs also resemble those (in pump power) found in the experiment. The fractal dimension of the set of gaps calculated from the model is $0.873 \pm 0.09$, which is closer to the value expected from a complete devil's staircase than the experimentally calculated value because the step in gain saturation energy can be made arbitrarily small in the model. Figure 6(c) illustrates the build-up phase of frequency locking, which again shows good agreement with the experimental results (Fig. 5(e)). As mentioned above, a small change in the initial polarisation state of the laser can trigger Farey fractions belonging to a different part of the Farey tree. This experimental observation is confirmed by the results shown in Supplementary Fig. 6, which have been obtained by slightly changing the linear intracavity loss in the model.

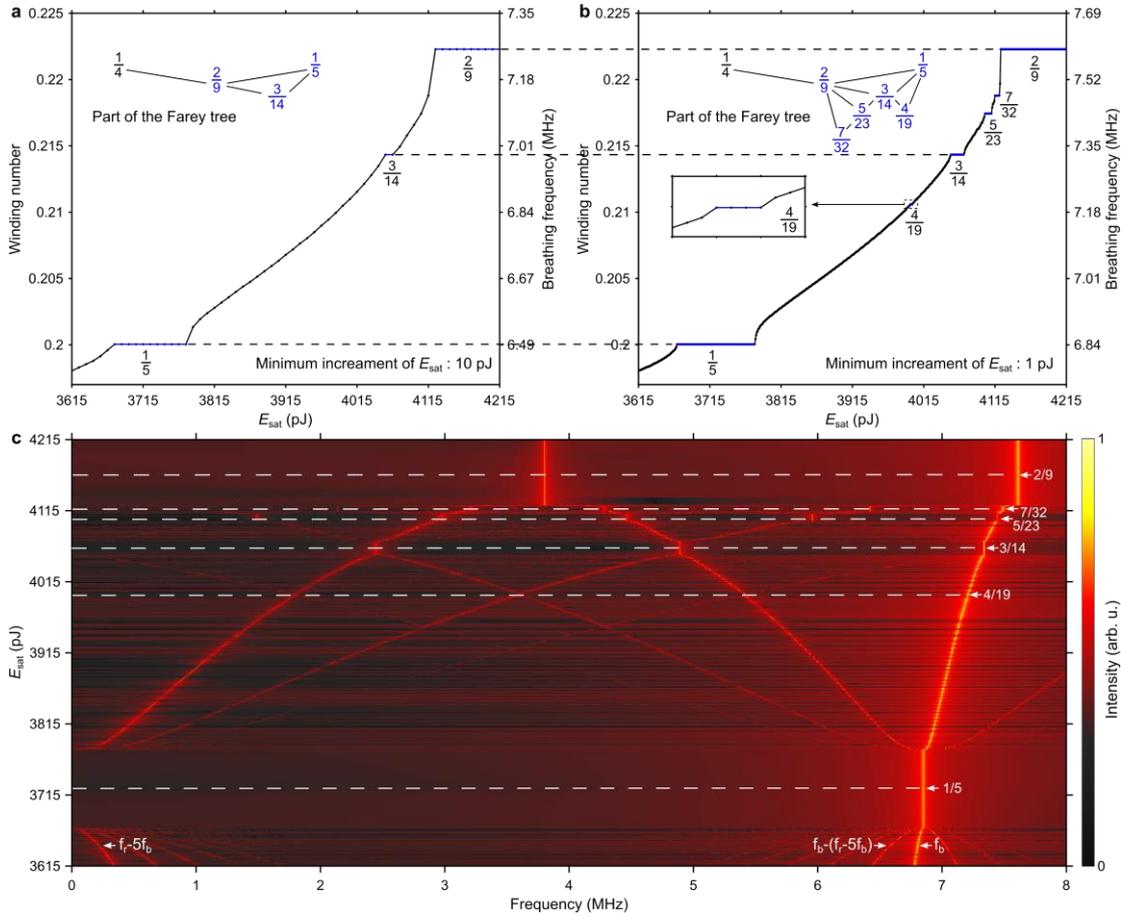

Fig. 6. Farey tree and devil's staircase observed in the numerical simulations. (a, b) Breathing frequency (winding number) as a function of the gain saturation energy (related to the pump power in the experiment) varied with a step of 10 and 1 pJ, respectively. With the smaller step, more plateaux are observed, evidencing a fractal pattern. In the insets are shown the parts of the Farey tree containing the observed Farey fractions. Since the plateau representing winding number 4/19 is very narrow, it is magnified in the inset in panel (b). (c) Map of spectral intensity in the space of radiofrequency and gain saturation energy, showing the build-up of rational winding numbers.

**Discussion**

We have demonstrated for the first time that a fibre laser working in the breathing-soliton generation regime is a nonlinear system showing frequency locking at Farey fractions. The frequency-locked breather states of the laser are characterised by robustness against parameter (pump power and polarisation) variations and a breathing frequency with narrow linewidth and high SNR. We have exploited the latter feature to realise intelligent control of the frequency-locking process, where the use of an EA with a locked breather-tailored merit function has been the key to the precise excitation of these breather states. Indeed, contrary to previous frequency-locking demonstrations in optics relying on an external modulation applied to the system, we have been able to manipulate the intrinsic breathing frequency of the laser system. The dimension of 0.906 determined from the measured devil's staircase indicates the universal nature of this nonlinear system. The breather mode-locked fibre laser thus may serve as a simple model system for the investigation of universal nonlinear properties. Besides, our work may stimulate the study of frequency locking in other physical systems where breathing solitons are found, where frequency locking could give a new angle on the dynamics of these systems. The EA approach used in this paper could benefit the control of the frequency-locking process in such systems as well as in others. We also believe that our EA-based approach for the control of frequency locking in fibre lasers is not restricted to NPE-based configurations and can be extended to other laser mode-locking schemes that entail period multiplication, such as the Mamyshev oscillator[51, 52].

Optical breathing solitons have been extensively studied in open-loop nonlinear systems such as, for example, single-pass fibre systems[36, 53]. However, in the absence of a frequency-locking mechanism, these breathers may suffer from instabilities originating from the noise of the input light. By contrast, we have studied the dynamics of breathers in a closed-loop system – a laser resonator. In this system, the universal frequency-locking process is tailored through the nonlinear interaction between the cavity repetition frequency provided by the laser resonator and the breathing frequency. Ergo, frequency-locked breathers can be generated, showing excellent stability against cavity parameter perturbations.

Frequency-locked breathers give rise to wide and dense RF combs which are not constrained by the length of the laser cavity. We have shown an example of a comb where the density of the spectral lines is increased by a factor of 41 compared with stationary single-pulse mode locking, thus enabling a line spacing in the sub-MHz range. Another example of a dense comb (where the density increase factor is 35) is given in Supplementary Fig. 4. Therefore, representing an alternative to fibre cavities of hundreds of meters which are regarded as being highly unstable, controlled frequency-locked breather lasers are attractive for many applications, for instance, in high-resolution spectroscopy.

We note that subharmonic entrainment of breather oscillations to the cavity repetition rate in a fibre laser was recently reported and explained as arising between the exceptional points of a non-Hermitian system involving two coupled modes with different detunings[26]. However, in light of the results presented in this paper, we believe that the observed dynamics of subharmonically entrained breathers fall outside of the exceptional point physics and can be well understood in the framework of resonances of a nonlinear system with two competing frequencies.

Dispersion plays an important role in determining the pulse dynamics in ultrafast fibre lasers[54, 55, 56]. The laser cavity used in this work has a nearly zero net dispersion. We have observed that frequency locking of breathers does not occur when the laser is operated at moderate or large normal dispersion[24, 49]. Thus, a very small net cavity dispersion seems to be crucial to the emergence of frequency-locked breathers in a fibre laser. It is worth noting in this regard that breathing solitons at nearly zero net dispersion and at large normal dispersion differ quite significantly in respect to their period of oscillation. Indeed, the former oscillate with a period ranging from several to dozens of round trips while the latter generally feature a much longer period of the order of hundreds of round trips[24], indicating that the underlying formation mechanism could be different. Future work will thoroughly investigate the connection between the frequency locking mechanism and the cavity dispersion.

## Methods

**Farey tree.** The Farey tree represents a particular ordering of the rational numbers by applying the Farey-sum or median operation $\oplus$ to two neighboring fractions, $m/n$ and $p/q$, which gives a new fraction in the next lower level of the tree by adding the numerators and the denominators separately: $\frac{m}{n} \oplus \frac{p}{q} = \frac{m+p}{n+q}$. The physically motivated hypothesis invoked to explain the local ordering of the hierarchy of (two-frequency) resonances is that the larger the denominator, the smaller the plateau. The Farey fraction or Farey mediant is the fraction with smallest denominator between $m/n$ and $p/q$, if they are sufficiently close that $|np - mq| = 1$ – when they are called adjacents – hence it is the most important resonance in the interval. The Farey tree provides a qualitative local ordering of two-frequency resonances and gives rise to a curve with an infinite number of steps showing self-similarity, which is known as the devil's staircase. For a detailed review see, e.g., Ref.[10].

**Fractal dimension of complementary set.** We have employed the equation[57]

$$\sum_i (S_i/S)^D = 1 \qquad (1)$$

for computation of the fractal dimension $D$ of the Cantor set complementary (on the pump-power axis) to a complete devil's staircase. In this equation, $S$ refers to the gap (in pump power) between two parental stairs representing winding numbers $m/n$ and $p/q$, and the $S_i$ correspond to the gaps between the filial stair $(m+p)/(n+q)$ and the parental stairs.

**Evolutionary algorithm.** The principle of the EA, as illustrated in Fig. 1(b), mimics mechanisms inspired by Darwin's theory of evolution: individuals composing a population progress through successive generations only if they are among the fittest[58]. Here, an individual refers to a laser regime, associated with the nonlinear transfer function determined by the three control voltages applied on the LCs; these voltages are therefore the genes of the individuals. The process begins with a collection of individuals or 'population', each comprising a set of randomly assigned genes. The system output is measured for each individual in the generation, evaluated by a user-defined merit function and assigned a score. The EA then creates the next generation by breeding individuals from the preceding generation, with the probability that an individual is selected to be a 'parent' based on their score ('roulette wheel' selection[58]). Two new individuals – children - are created from the crossover of two randomly selected parents, namely the interchange of their genes. A mutation probability is also specified, allowing for the genetic sequence to be refreshed. This process repeats until the algorithm converges and an optimal individual is produced. In the experiment, the algorithm is initialised with a population of 50 individuals and the population size of the next generations is kept constant to 30 individuals (6 parents and 24 children). Evaluation of the properties of an entire generation of individuals typically takes 2.5 minutes.

A critical factor to the success of a self-optimising laser implementation is the merit function, which must return a higher value when the laser is operating closer to the target regime. In the present work, we have defined and tested the following merit function for the auto-setting of an optimised self-starting frequency-locked breather regime:

$$F_{\text{merit}} = \alpha F_{\text{ml}} + \beta F_{\text{b}} + \gamma F_{\text{snr}} \qquad (2)$$

In Eq. (2), $F_{\text{ml}} = \sum_{i=1}^{i=N} I_i / L, I_i = \begin{cases} I_i, (I_i \geq I_{\text{th}}) \\ 0, (I_i < I_{\text{th}}) \end{cases}$ is the merit function relating to the mode-locked laser operation[59], where $N$ is the number of laser output intensity points recorded by the oscilloscope ($N=2^{24}$, corresponding to a time trace of 7174 cavity round-trips), $I_i$ is the intensity at point $i$ and $I_{\text{th}}$ is a threshold intensity that noise should not exceed. Thus, $F_{\text{ml}}$ represents the average of pulses' intensities, and is used to exclude laser modes, such as relaxation oscillations and noise-like pulse emission, which may display similar RF spectral features to the breather regime. The second term $F_b$ is a merit function that discriminates between breather and stationary pulsed operations, derived from the feature that the breathing frequency $f_b$ manifests itself as two symmetrical sidebands $f_{\pm 1}$ around the cavity repetition frequency $f_r$ in the RF spectrum of the laser output ($f_b = |f_{\pm 1} - f_r|$).There are no sidebands when the laser works in a stationary mode-locking regime. Therefore, $F_b$ is designed to exploit the intensity ratio of the central band located at $f_r$ to the sidebands at $f_{\pm 1}$: $F_b = 1 - \sum_{f=f_r-\Delta}^{f=f_r+\Delta} I(f) / \sum_{f=f_{-1}}^{f=f_{+1}} I(f)$, where the numerator and denominator in the fraction are the intensities measured across the width $2\Delta$ of the frequency band centred on $f_r$ and the frequency interval from $f_{-1}$ to $f_{+1}$, respectively. Accordingly, if $F_b$ approaches zero, it means that $f_r$ prevails and there are no sidebands, indicating a stationary mode locking state. On the contrary, $F_b$ values far from zero evidence the presence of strong sidebands in the RF spectrum, indicating possible breather formation in the laser cavity. In the optimisation procedure, the RF spectrum is obtained directly from the oscilloscope that processes the fast Fourier transform of the laser output intensity recording. The weighted sum of $F_{ml}$ and $F_b$ can be used for the self-optimisation of the breather mode locking regime[49]. The third term in Eq. (2) is a new merit function that discriminates between frequency-locked and unlocked breather oscillations by evaluating the strongest breathing frequency in the interval $[f_r + \delta, 3f_r/2]$ : $F_{snr} = \max I(f), f \in [\,f_r + \delta, 3f_r/2]$ , where the frequency shift $\delta$ is used to exclude the fundamental frequency from the evaluation interval, and $f_r/2$ represents the maximum possible breathing frequency. The weights of the three components in Eq. (2) are determined empirically and set to $\alpha$=2000, $\beta$=200 and $\gamma$=200.

**Numerical modelling.** The pulse propagation in the optical fibres is modelled by a generalised nonlinear Schrödinger equation, in the scalar approach, which takes the following form[50]:

$$\psi_z = -\frac{i\beta_2}{2}\psi_{tt} + i\gamma|\psi|^2\psi + \frac{g}{2}\left(\psi + \frac{1}{\Omega^2}\psi_{tt}\right), \qquad (3)$$

where $\psi = \psi(z,t)$ is the slowly varying electric field moving at the group velocity along the propagation coordinate $z$, $\beta_2$ and $\gamma$ are the second-order dispersion and Kerr nonlinearity coefficients, respectively, and the dissipative terms represent linear gain as well as a parabolic approximation to the gain profile with the bandwidth $\Omega$. The gain is saturated according to $g(z) = g_0 \exp(-E_p/E_{sat})$, where $g_0$ is the small-signal gain, which is non-zero only for the gain fibre, $E_p(z) = \int dt |\psi|^2$ is the pulse energy, and $E_{sat}$ is the gain saturation energy determined by the pump power. The effective nonlinear saturation involved in the NPE mode-locking technique is modelled by an instantaneous and monotonous nonlinear transfer function for the field amplitude: $T = \sqrt{1 - q_0 + q_m/[1 + P(t)/P_{sat}]}$, where $q_0$ is the unsaturated loss due to the absorber, $q_m$ is the saturable loss (modulation depth), $P(z,t) = |\psi(z,t)|^2$ is the instantaneous pulse power, and $P_{sat}$ is the saturation power. Linear losses are imposed after the passive fibre segments, which summarise intrinsic losses and output coupling. The numerical model is solved with a standard symmetric split-step propagation algorithm and using similar parameters to the nominal or estimated experimental values (see Supplementary Table 1).

**Data availability**

The data that support the findings of this study are available from the corresponding author upon reasonable request.

**Code availability**

The code that support the findings of this study are available from the corresponding author on request.

**Acknowledgements**

We acknowledge support from the National Natural Science Fund of China (11621404, 11561121003, 11727812, 61775059, 12074122, 62022033 and 11704123), Shanghai Municipal Science and Technology Major Project (2019SHZDZX01-ZX05), Key Project of Shanghai Education Commission (2017-01-07-00-05-E00021), National Key Laboratory Foundation of China (6142411196307), Shanghai Rising-Star Program and Science and Technology Innovation Program of Basic Science Foundation of Shanghai (18JC1412000), the Agence Nationale de la Recherche (ANR-20-CE30-004)

**Author contributions**

J.P. and H.Z. initiated the work. X.W., J.P. and H.Z. performed the experiments. X.W., Y.Z., J.P. and S.B. carried out the numerical simulations. J.P. S.B. C.F and H.Z. supervised and guided the work. All authors contribute to data analysis and the writing of the paper.

**Competing interests**

The Authors declare no Competing Financial or Non-Financial Interests.